**On the transferability of three water models developed by adaptive force matching**

Hongyi Hu, Zhonghua Ma, and Feng Wang*

Department of Chemistry and Biochemistry,

University of Arkansas,

Fayetteville,

Arkansas, 72701,

United States

* Corresponding Author, Email: fengwang@uark.edu, Phone: (479)-575-5625, Fax: (479)-575-4049

**Abstract**

Water is perhaps the most simulated liquid. Recently three water models have been developed following the adaptive force matching (AFM) method that provides excellent predictions of water properties with only electronic structure information as a reference. Compared to many other electronic structure based force fields that rely on fairly sophisticated energy expressions, the AFM water models use point-charge based energy expressions that are supported by most popular molecular dynamics packages. An outstanding question regarding simple force fields is whether such force fields provide reasonable transferability outside of their conditions of parameterization. A survey of three AFM water models, B3LYPD-4F, BLYPSP-4F, and WAIL are provided for simulations under conditions ranging from the melting point up to the critical point. By including ice-Ih configurations in the training set, the WAIL potential predicts the melting temperate, $T_M$, of ice-Ih correctly. Without training for ice, BLYPSP-4F underestimates $T_M$ by about 15 K. Interestingly, the B3LYPD-4F model gives a $T_M$ 14 K too high. The overestimation of $T_M$ by B3LYPD-4F mostly likely reflects a deficiency of the B3LYP reference. The BLYPSP-4F model gives the best estimate of the boiling temperature $T_B$ and is arguably the best potential for simulating water in the temperature range from $T_M$ to $T_B$. None of the three AFM potentials provides a good description of the critical point. Although the B3LYPD-4F model gives the correct critical temperature $T_C$ and critical density $\rho_C$, there are good reasons to believe the agreement is reached fortuitously. Links to Gromacs input files for the three water models are provided at the end of the paper.

**I. Introduction**

Water is ubiquitous and generally considered to be one of the most versatile liquids. It is not surprising that a significant amount of simulation has been done to investigate various properties of water. There are probably more potentials developed for water than any other liquid. [1] Early models that gained significant popularity include SPC[2], SPC/E[3], TIP3P[4], and TIP4P[5]. All of these models were developed by fitting to experimental properties. Of these, arguably, SPC/E and TIP4P are considered to be the most successful. In recent years, new members in the TIP4P family, such as TIP4P-Ew[6] and TIP4P-2005[7], have been created and are generally believed to be more accurate than the earlier ones.

Although maybe not as popular as experiment based potentials, quite a few water models were developed by fitting to electronic structure calculations. Early electronic structure based models, such as MCY[8], fail to predict several key properties, such as the density of water. Consequently, these models are not as widely used as experiment based potentials. In recent years, electronic structure based potentials, such as the TTM family [9, 10], DPP family [11, 12], HBB family [13, 14], and many others,[15, 16] are more sophisticated and accurate. However, these potentials are rather expensive to evaluate and have only limited support in public domain molecular dynamics (MD) packages.

Although experiment based potentials satisfactorily reproduce the most important properties, it is very hard to judge if a property is reproduced for the correct reason. Also, it is hard to determine if such potentials can reliably predict properties not being fit. In this sense, a potential fit only to electronic structure information is more robust in that if such a potential does reproduce an experimental property, it is more likely that such an agreement is obtained by correctly capturing the underlying physics.

Recently, several water potentials were developed based on the adaptive force matching (AFM) approach.[17-21] These water models were created by only fitting to electronic structure calculations. With AFM, the fit was done iteratively in the condensed phase. Obtaining reference forces in the condensed phase allows fitting of relative simple energy expressions that implicitly capture many-body effects. Only energy expressions that are supported by popular MD packages, such as Gromacs, were used in typical force fields developed by AFM[20-23]. With simple point-charge based energy expressions, the three water models investigated in this work require computational resources comparable to that of TIP4P for each force evaluation. It is worth mentioning that these models are generally a factor of two slower than TIP4P due to the requirement for smaller time steps. However, the use deuterium to replace hydrogen in simulations alleviates this disadvantage.

Although simple energy expressions lead to efficient force fields, they may limit the transferability of the potential. The philosophy of AFM is to fit a force field for a specific condition. This is achieved by including, in the training set, only reference configurations representative of the condition of interest. This is actually not very different from the development of some, if not most, experimental based force fields, where only experimental properties under limited conditions were fit. For example, the TIP4P potential was fit only to properties at 1 atm and 25°C.[5] Nonetheless, these water models are frequently used under thermodynamic conditions not tested during parameterization.[24-27]

Several water models have been created based on AFM. Some of the models [20, 21] were designed to be used with *ab initio* free energy perturbation theory.[28, 29] Three recent water models, B3LYPD-3F, BLYPSP-4F, and WAIL, offer similar performance and are capable of simulating the liquid states[17-19]The objective of this paper is to investigate the performance of these three models outside of the thermodynamic conditions of parameterization. Under the conditions of parameterization, the AFM models have been found to be highly competitive with experiment based potentials. For example, the WAIL potential designed for the modeling of ice and water gives a very good description of the melting temperature, $T_M$, of ice and temperature of maximum density, TMD, of water. It is interesting to see if these models are better or worse than experimental based potentials outside their "comfort zone". These results should establish the applicability of these models as general purpose potentials for water.

In order to accomplish this purpose, we investigate $T_M$, diffusion constant (D), viscosity ($\eta$), surface tension ($\gamma$), static dielectric constant ($\varepsilon_s$), TMD, boiling temperature ($T_B$), critical temperature ($T_C$), critical density ($\rho_C$), and critical pressure ($P_C$). In this paper, we will provide a brief review of the AFM procedure in section II, and a brief summary of the three water models in section III. Computational details will be reported in section IV. Results and summary are presented in section V. A conclusion will be included as section VI.

**II. The adaptive force fitting procedure**

AFM was designed to fit a force field to best reproduce electronic structure forces obtained under a particular thermodynamic condition or a set of thermodynamic conditions of interest. AFM requires an initial guess to the force field. From such a force field, a typical realization of AFM contains three steps as illustrated in Fig. 1.

The first step in AFM is the sampling step. In this step, the phase space associated with thermodynamic conditions of interest is traversed with a sampling algorithm, such as MD or Monte Carlo (MC). The guess force field will be used to integrate MD or MC trajectories. Configurations are randomly selected from the trajectories to form the training set. Standard sampling algorithms traverse the phase space according to the Boltzmann weight of each microstate. More important regions of the phase space are thus better represented in the training set. It is also straight-forward to couple the sampling algorithm with techniques that facilitate barrier crossing for more challenging systems.

The second step of AFM is the *ab initio* step. In this step, an electronic structure method is used to obtain reference forces. If the system size is too large to afford an adequate quantum mechanics (QM) treatment, QM/molecular mechanics (MM) modeling can be used by treating only part of the system with an electronic structure method. The MM region will also be described by the guess force field. Through Coulombic embedding, the more extended MM region allows the QM part to feel an environment similar to that in the condensed phase.

The last step of AFM is the fitting step. In this step, the electronic structure forces obtained in the *ab initio* step are used to reparameterize the force field. Non-linear optimization is a grand challenge. Fortunately, many parameters in popular force field energy expressions are linear parameters with respect to forces. We rely on a mixed optimization procedure, where the linear parameters are determined with the singular value decomposition (SVD) method and the number of non-linear parameters is kept at a minimum. SVD is very robust for optimizations that involve only linear parameters.

With the more accurate force field obtained in the fitting step as the guess force field, AFM restarts from the sampling step. The improved force field will lead to a better training set and a more realistic representation of the MM environment for the QM/MM calculations. AFM terminates when the force field parameters no longer changes appreciably over a few generations. . At this point, all the QM forces from all the converged generations will be fit together in a global fit to reduce the error bar on the final parameters.

Only forces are used with AFM. Comparing with an energy matching method, the use of forces has two advantages. For a system containing $N$ atoms, $3N$-6 force values are typically available for fitting, whereas there is only one total energy. For an electronic method with analytical derivatives implemented, the calculation of forces is typically no more than a few times the cost of calculating energy. For example, with density functional theory (DFT), the majority of the CPU time is spent iterating the density matrix to convergence. Once self-consistency has been reached, the determination of forces does not require much additional CPU cycles. Even for correlated electron methods, such as MP2, the self-consistent field (SCF) step can take a significant portion of computational resource, especially when the QM region is small, *e. g.* with around 100 valence electrons for MP2. It is thus advantageous to use forces for more efficient fitting.

A second advantage of force fitting is the convenience in removing boundary effects when used with QM/MM. With energy matching, the influence of the QM/MM boundary to the total energy is hard to quantify. On the other hand, it is straightforward to remove boundary effects by discarding forces on boundary atoms in AFM[17]. The boundary forces are less accurate since these forces are influenced by nearby MM atoms. Only forces on atoms buried in the QM region are used in AFM.

**III. Three water models developed by AFM.**

In this work, we survey the properties of three water models, BLYPSP-4F, B3LYPD-4F, and WAIL.[17-19] Both BLYPSP-4F and B3LYPD-4F were created for simulation of liquid water in the temperature range from 0 to 40 °C. The WAIL potential was developed with both ice and liquid water configurations in the training set. The WAIL potential was designed to simulate ice-water equilibrium around 273 K under 1 atm.

Both the WAIL model and the BLYPSP-4F potential were developed using the BLYP-SP method for obtaining reference forces.[30, 31] Thus the only difference between them is the different training set. The BLYP-SP method was trained by the quadratic configuration interaction with single and double excitations (QCISD) method, which gives forces almost identical to those from the coupled cluster with single and double excitations (CCSD) method.[31] This is expected because QCISD is a simplified case of CCSD.[32-34] It is safe to assume that the WAIL and BLYPSP-4F force fields were fit indirectly to a coupled cluster quality potential energy surface. The B3LYPD-4F model was fit to reference forces calculated with the B3LYP exchange functional with an additional dispersion term. The dispersion term was determined by symmetry adapted perturbation theory (SAPT) [35] calculations performed on gas phase dimers.[17]

All three force fields share the same energy expression that was optimized for water with AFM.[17] The total configuration energy includes an intra-molecular part with the form,

$$U_{\text{intra}}(r_1, r_2, \theta) = \frac{k_2}{2}(r_1 - r_e)^2 + \frac{k_3}{3}(r_1 - r_e)^3 + \frac{k_4}{4}(r_1 - r_e)^4 \\ + \frac{k_2}{2}(r_2 - r_e)^2 + \frac{k_3}{3}(r_2 - r_e)^3 + \frac{k_4}{4}(r_2 - r_e)^4 \\ + \frac{k_\theta}{2}(\theta - \theta_e)^2 \quad , \qquad (1)$$

and an inter-molecular part with the form,

$$U_{\text{inter}}^{ij, i \neq j} = A_{OO} \exp(-\alpha r_{O_i O_j}) - f(r_{O_i O_j}) \frac{C_{OO}}{r_{O_i O_j}^6} \\ + \sum_{\mu \in i} \sum_{\upsilon \in j} \frac{q_\mu q_\upsilon}{r_{\mu\upsilon}} + \sum_{\mu \in i} U_{\text{HB}}(r_{M_j H_\mu}) \quad , \qquad (2)$$

where $i$ and $j$ are molecular indices; $u$ and $v$ are atomic indices.

For the intramolecular term, $U_{intra}$, $k_2$, $k_3$, $k_4$ and equilibrium bond length $r_e$ are parameters for the quartic bond term, $k_\theta$ is bond constant for the harmonic bond term and $\theta_e$ is the equilibrium HOH angle. $r_1$, $r_2$ and $\theta$ are the OH bond lengths and the HOH bond angle, respectively.

The first two terms of the intermolecular contribution $U_{inter}$ is similar to a Buckingham potential with $f(r_{O_iO_j})$ being the Fermi damping function for the B3LYPD-4F model[36] and being a constant of one for the other two models. The third term is intermolecular Coulombic interactions and the last term,

$$U_{HB}(r_{MH}) = \begin{cases} A_4/r_{MH}^4 - A_4/r_c^4 + 4(r_{MH}-r_c)/r_c^5 & r \leq r_c \\ 0 & r > r_c \end{cases}, \quad (3)$$

is a short-range repulsion term introduced to improve the description of hydrogen bonds for a point-charge model.[37] In Equation 3, $r_{MH}$ is the intermolecular M-H distance and $r_c$ is a cut-off to keep $U_{HB}$ short-range. All three models place the negative charge on the $M$ site defined by

$$\vec{r}_{OM} = a(\vec{r}_{OH_1} + \vec{r}_{OH_2}), \quad (4)$$

where $\vec{r}_{OM}$ is the location of the M site relative to the oxygen and $a$ is chosen to be 0.20.

All the parameters of the three models are summarized in Table 1. The accuracy of the models has been tested under the conditions they were parameterized for. In summary, with nuclear quantum effects treated with path integral MD, all the models faithfully reproduce the radial distribution functions (RDF), the heat of vaporization, $\Delta H_{vap}$, and the heat capacity of water. In addition, the densities of these models are also in good agreement with experiment. Although densities were only calculated with classical MD, the nuclear quantum effects are only expected to influence water density slightly.

One way to estimate to what extent nuclear quantum effects are expected to influence a thermodynamic property is to compare the experimental values for $H_2O$ and $D_2O$. In this context, a thermodynamic property is defined as a property that can be calculated from the system partition function. Within classical statistical mechanics, the momentum contribution to the partition functions can be separated and integrated out analytically. Regardless of particle masses, the ideal gas partition function can always be recovered for the momentum degrees of freedom. On the other hand, the potential energy contribution to the partition function is not affected by particle masses within the Born-Oppenheimer approximation. Thus, classical statistical mechanics predicts thermodynamic properties to be insensitive to isotope masses. $H_2O$ and $D_2O$ are thus expected to have the same number density $\rho_n$. At 293 K, the experimental $\rho_n$ for $H_2O$ and $D_2O$ are 55.41 mol/L and 55.19 mol/L, respectively.[38] The small difference is caused by nuclear quantum effects.

## IV. Computational details

Since the vast majority of MD simulations were performed within the framework of classical mechanics, we investigate the transferability of the AFM water models without taking nuclear quantum effects into consideration. The equation of motion is integrated with a leap-frog integrator with a 0.5 fs time step. The hydrogen mass was chosen to be 1.008 g/mol for all the simulations except for the investigation of liquid-vapor critical point at elevated temperatures. For these trajectories, the hydrogen atoms were replaced with deuterium for improved stability without the need for reduced timestep sizes that are occasionally associated with elevated temperatures. The long-range electrostatic interactions are treated with the particle mesh Ewald method with an Ewald precision of $10^{-6}$.

The bulk properties of liquid, such as shear viscosity ($\eta$), diffusion constant ($D$), dielectric constant and density were measured at 300 K in an orthorhombic box containing 1728 water molecules. The box has a typical dimension of 3.60 nm by 3.60 nm by 4.00 nm with a small variation due to the different equilibrium density for each water model. To determine $\eta$, the periodic perturbation method[39] as implemented in Gromacs was used. Five 3ns NVT trajectories were integrated with the last 1 ns of each used for the actual determination of $\eta$. For the calculation of $D$, 10 ns of trajectories were integrated with the mean square displacement (MSD) calculated over the last 5 ns. $D$ was extracted using the Einstein relation by fitting the MSD from 10 to 50 ps.

The dielectric constant of water was calculated with the fluctuation and dissipation theorem with the formula[40]

$$\varepsilon_{rv} = \varepsilon_{\infty} + \frac{4\pi}{3Vk_BT}\left(\langle M^2 \rangle - \langle M \rangle^2\right), \tag{5}$$

where a value of 1 is used for $\varepsilon_{\infty}$ and $M$ is the dipole moment of the simulation box. In Eq. 5, $V$, $k_b$, and $T$ are the volume, Boltzmann constant, and the temperature, respectively. Since none of the water models have explicit treatment of polarizability, the dielectric constant in Eq. 5 can only arise from molecular rotation and vibration. Thus, the subscript $rv$ is included as a reminder. Further discussion is provided in Sec. V about the relationship between $\varepsilon_{rv}$ and the statistic dielectric constant $\varepsilon_s$.

The surface tension calculations were performed with a 1728 water slab in an orthorhombic box of 3.60 nm by 3.60 nm by 8.00 nm. The slab normal is along the long axis of the box. The slab has a thickness of approximately 4 nm leaving a vacuum region of approximately 4 nm in depth. $\gamma$ was calculated with the formula

$$\gamma = \frac{1}{2}(P_Z - \frac{P_X + P_Y}{2}), \qquad (6)$$

where $P_X$, $P_Y$, and $P_Z$ are the diagonal elements of the system stress tension and the prefactor ½ is due to the slab having two surfaces. For each trajectory, 10 ns of NVT simulations were performed with the last 5 ns used for measuring the pressure. The final results are averaged over five trajectories simulated with uncorrelated initial configurations.

The $T_M$ of ice-Ih was estimated with the direct coexistence method [41, 42] by monitoring the stability of an ice-water interface. During the direct coexistence simulation, the system pressure was kept at 1 bar with the Parrinello-Rahman barostat with a relaxation time of 5 ps and the temperature was controlled by the Nosé-Hoover thermostat with a relaxation time of 0.2 ps.

The critical properties were studied by slab simulations at temperatures from 0.70 to 0.95 $T_c$. A 4.0 × 4.0 × 6.0 nm$^3$ orthorhombic box containing 1048 water molecules was used for the critical point calculations. A liquid and vapor interface can be clearly identified in these simulations. The $T_c$ was determined with the Wegner expansion[43] with

$$\rho_l - \rho_g = A_0|\tau|^{\beta_c} + A_1|\tau|^{\beta_c+\Delta} + A_2|\tau|^{\beta_c+2\Delta} + A_3|\tau|^{\beta_c+3\Delta}, \quad (7)$$

where $\tau=1-T/T_c$, $\beta_c = 0.325$, and $\Delta = 0.5$. $A_0$, $A_1$, $A_2$, $A_3$ are parameters to be fit. These exponents were determined with renormalization group theory.[27]

The $\rho_C$ was obtained with[44]

$$\rho_l + \rho_g = 2\rho_c + D_{1-\alpha}|\tau|^{1-\alpha} + D_1|\tau|, \qquad (8)$$

where an α of 0.11 is used; $D_{1-\alpha}$, and $D_1$ are parameters to be fit. The relationship between the equilibrium temperature and pressure were fitted using the Antoine's law

$$\ln(P) = A + \frac{B}{T+C}. \tag{9}$$

where A, B, C are parameters to be fit. With the Antoine's law, $P_C$ can be obtained by setting the temperature to $T_C$. Similarly, $T_B$ was determined with the Antoine's law by solving for the temperature that results in a pressure of 1 atm. It is worth noting that the pressure that goes in Eq. 9 is the pressure normal to the liquid-vapor interface. The stress in the liquid plane has an additional contribution from the surface tension. Our procedure for determining the $T_B$, $T_C$, $\rho_C$, and $P_C$ was validated by fitting to the TIP4P density and vapor pressure reported by Vega.[45]

**V. Results and discussion**

The properties of the three models and those from TIP4P/2005 were reported in Table 2 along with the corresponding experimental values for $H_2O$ and $D_2O$. For all thermodynamic properties, the difference between $H_2O$ and $D_2O$ gives a rough estimate of the importance of nuclear quantum effects. Since a classical simulation is used when calculating the properties, it is more appropriate to compare thermodynamic properties with the corresponding $D_2O$ values. In Table 2, only $D$ and $\eta$ are not thermodynamic properties.

Rather than listing the mass density, the number density ($\rho_n$) is reported in Table 2. As mentioned previously, classical statistical mechanics require $\rho_n$ to be identical for $H_2O$ and $D_2O$. All three AFM models predict $\rho_n$ to be within 2% of the experimental value for $D_2O$. The TIP4P/2005 models give the best estimate of $\rho_n$. This is expected because density at the ambient condition is a fitting parameter for the development of the TIP4P/2005 potential.[7]

The WAIL model was parameterized with both ice-Ih and liquid water configurations in the training set. The WAIL model predicts the ice-Ih $T_M$ to be 270 K,[46] in good agreement with experiments. The BLYPSP-4F model uses an energy expression identical to that used by WAIL and was fit to reference forces calculated with the same method. The only difference between the two models is the absence of ice-Ih configurations in the parameterization of BLYPSP-4F. The BLYPSP-4F model predicts an ice-Ih $T_M$ of 258 K, which is 15 K below the experimental value for $H_2O$. This is expected since BLYPSP-4F was designed to give the best fit only for liquid water, thus the model predicts ice-Ih to be less stable than it actually is. A reduced stability of the solid phase results in a lower $T_M$.

It is most interesting that B3LYPD-4F has a $T_M$ of 287 K, significantly above that of the experimental value. The B3LYPD-4F model was created by force matching the B3LYP reference forces with additional SAPT based dispersion. It has been established by Xantheas et al. that popular DFT functionals overestimates the $T_M$ of ice-Ih. For example, the PBE functional gives an ice-Ih $T_M$ of 417 K and BLYP gives a $T_M$ of 411K.[47] Even with dispersion correction, BLYP still overestimates $T_M$. With Grimme's dispersion correction,[48-50] BLYP-D ice-Ih has a $T_M$ of 360 K.[51] Due to the mixing of exact Hartree-Fock exchange, B3LYP is much more computationally intensive than BLYP. The B3LYP $T_M$ for ice-Ih is unknown. Although B3LYP is generally believed to be more accurate than BLYP, it is very likely B3LYP also over-estimates $T_M$ even with dispersion correction. The higher $T_M$ of the B3LYPD-4F ice most likely suggests that the B3LYP functional does over-estimate $T_M$ even with dispersion correction. If this statement is true, it seems to indicate that force fields created with AFM do reflect the underlying physics described by the method used to create reference forces. The high $T_M$ is also consistent with the B3LYPD-4F potential giving too large a $\Delta H_{vap}$ and too low a $D$. Even with path integral, B3LYPD-4F overestimates $\Delta H_{vap}$ by 1 kcal/mol.

Of the three models, BLYPSP-4F gives the best $D$ when compared with experiments. Although this is from a classical simulation, if centroid molecular dynamics[52-55] is used to calculate $D$, the correction due to nuclear quantum effects is expected to be small.[19, 56] This is a result of the quartic bond term used to describe OH stretch in these models. Nuclear quantum effects are expected to increase water dipole moment with such an anharmonic bond term. The increased dipole moment causes $D$ to decrease, which act in the opposite direction of quantum facilitated barrier crossing.

Although the BLYPSP-4F model gives the best $D$, it slightly underestimates $\eta$. Of the three AFM models, the B3LYPD-4F most seriously overestimates water viscosity, consistent with its higher $T_M$. The $\eta$ of both BLYPSP-4F and WAIL can be considered satisfactory. If the Stokes-Einstein relation for spherical particle is applicable to liquid water, the Stokes radii are 1.20 Å, 1.26 Å, and 1.45 Å for the B3LYPD-4F, BLYPSP-4F, and WAIL models, respectively.

All three AFM water models give satisfactory description of water surface tension with the best agreement produced by B3LYPD-4F. The WAIL potential also gives a good $\gamma$ with an error slightly more than 5 %.

All three AFM models significantly underestimate the dielectric constant of water when Eq. 5 is used. However, this is expected since none of the three models account for polarization effect explicitly. Electronic polarization is responsible for the high frequency component of water dielectric constant, which is approximately 1.78. Without explicit treatment of polarization, only the dielectric contribution due to rotation and vibration is captured by these force fields. This is calculated with Eq. 5 and represented by $\varepsilon_{rv}$ in the table. It has been argued that the optical contribution to the static dielectric constant ($\varepsilon_s$) can be accounted for by multiplying $\varepsilon_{rv}$ by the high frequency dielectric constant of 1.78.[57] This procedure gives a $\varepsilon_s$ in excellent agreement with experiments for all three models with the BLYPSP-4F model being the best and the WAIL potential equally good considering the error bar. With the 1.78 scaling, the TIP4P/2005 overestimates the $\varepsilon_s$ by 37 %.

The liquid TMD is approximately 10 K above $T_M$ for the BLYPSP-4F potential and 12 K above $T_M$ for the WAIL potential. While the $H_2O$ TMD is about 4 K above $T_M$, the $D_2O$ and $T_2O$ TMDs are 7 K and 9 K above their respective $T_M$. Although when comparing the absolute temperature, the WAIL potential is giving the best agreement with experimental TMD, if relative difference is of interest, the BLYPSP-4F model is likely to provide the best agreement. The B3LYPD-4F potential has a TMD about 18 K above its $T_M$. If the maximum density of water was indeed caused by a second critical point in the supercooled region, [58-62] a high TMD is likely to indicate a second critical point closer to the already elevated $T_M$. The B3LYPD-4F potential may be a good model for investigating the thermodynamics of water near its putative second critical point.

Figure 2 shows a fit to Antoine's equation for the three models. The $T_B$ under 1 atm and $P_C$ can be obtained from this figure. As mentioned in Sec. IV, $T_C$ and $\rho_c$ were determined by the Wegner expansion. All three models overestimate $T_B$ with the BLYPSP-4F model being the closest to the experimental value. The BLYPSP-4F $T_B$ is only 20 K higher than the experimental value. From Fig. 2 all three models underestimate the vapor pressure. This is not surprising considering these AFM force fields were only optimized for the condensed phases. If the force field parameters were optimized for the gas phase, one would anticipate the gas phase to be more stable thus leading to a higher vapor pressure and lower $T_B$.

The B3LYP-4F model actually gives excellent agreement with respect to both $T_C$ and $\rho_c$. This is most likely due to a fortuitous cancelation of errors considering the poor prediction of $P_C$ by this model  Both the BLYPSP-4F and the WAIL potentials over-estimates $T_C$ but underestimates $\rho_c$. All three AFM models underestimate $P_C$, with the B3LYP-4F model showing the largest error when compared with experiments.

Comparing the three AFM models, the BLYPSP-4F model predicts a $T_B$ of 120 C and a $T_C$ of 685 K is the most successful at elevated temperatures. The TIP4P/2005 model parameterized by fitting experimental properties performs better around the critical point. This may be due to the TIP4P/2005 model giving the correct $\Delta H_{vap}$ and density in a classical simulation by construction. None of the AFM models were fit to the experimental density. A 2% error in liquid density may lead to larger percentage errors in other properties. On the other hand, only properties from AFM models can be considered as a first-principle based prediction.

**VI. Conclusion.**

Several properties of liquid water ranging from the melting point to the critical point were calculated for three water potentials developed with the AFM method to assess the transferability of AFM potentials outside of the condition of parameterization. The three models, B3LYPD-4F, BLYPSP-4F, and WAIL share the same energy expressions and were fit only to high quality electronic structure calculations. While the B3LYPD-4F potential was fit to dispersion corrected B3LYP forces, the BLYPSP-4F and WAIL potentials were fit to coupled cluster quality forces obtained with the BLYP-SP method. Only liquid configurations were used in the parameterization of the B3LYPD-SP and the BLYPSP-4F potentials. Both ice and water configurations were included in the training set for the WAIL potential.

Including the ice configurations in the training set allows the WAIL potential to produce a good $T_M$ of 270 K. On the other hand, the $T_M$ of BLYPSP-4F is 15 K below the experimental value. A similar but even more serious underestimation is observed in many other water potentials created by fitting to experimental properties of the liquid. The B3LYPD-4F potential over-estimates the $T_M$ by 14 K. This most likely reflects a more significant overestimation of $T_M$ by the B3LYP reference method even with a SAPT based dispersion correction.

All of the water models studied significantly underestimate the vapor pressure of liquid water with the BLYPSP-4F model giving the best $T_B$ and the B3LYPD-4F model fortuitously giving the best $T_C$ and $\rho_c$. Several other properties such as $\gamma$, $\varepsilon_s$, and $\eta$ are calculated under 1bar and 300 K. At temperature below $T_B$, BLYPSP-4F is arguably the best model for liquid water. The experimental based TIP4P/2005 model is better than any of the AFM based models close to the critical point. However, TIP4P/2005 still seriously underestimates $P_C$. We believe that it is a good idea to use the BLYPSP-4F model for liquid simulations below $T_B$ and use the WAIL model for simulations involving ice and water.

Gromacs input files for all three water models, B3LYPD-4F, BLYPSP-4F, and WAIL can be downloaded at http://wanglab.uark.edu/ARCC_Wang_AFM.

**VII. Acknowledgement**

This work was supported by NSF CAREER award CHE0748628 and by the startup grant from University of Arkansas. The computer resources for this study were provided by the Arkansas High Performance Computational Center through grant MRI-R2 #0959124 provided by the NSF.

Table 1. Parameters for the B3LYPD-4F, BLYPSP-4F and WAIL water models previously developed with AFM.

|  | B3LYPD-4F | BLYPSP-4F | WAIL |
|---|---|---|---|
| $q_M$ (e) | -1.346 | -1.3290 | -1.373 |
| $q_H$ (e) | 0.673 | 0.6645 | 0.686 |
| $a$ | 0.2 | 0.2 | 0.2 |
| $A_{oo}$ ($10^3$ kcal/mol) | 267.412 | 210.710 | 201.3 |
| $\alpha$ (1/Å) | 4.25 | 4.055 | 3.98 |
| $C_{oo}$ (kcal·Å$^6$/mol) | 610.578 | 610.578 | 770 |
| $A_4$ (kcal·Å$^4$/mol) | 73.97 | 81.489 | 77.80 |
| $r_c$ (Å) | 2.483 | 2.483 | 2.483 |
| $r_e$ (Å) | 0.956 | 0.951 | 0.9496 |
| $k_2$ (kcal /(mol·Å$^2$)) | 1200.61 | 1255.19 | 1270 |
| $k_3$ (kcal /(mol·Å$^3$)) | -4427.34 | -4503.57 | -4860 |
| $k_4$ (kcal /(mol·Å$^4$)) | 8501.88 | 7020.31 | 10310 |
| $\theta_e$ (°) | 107.33 | 106.678 | 106.89 |
| $k_\theta$ (kcal/(mol·rad$^2$)) | 77.78 | 82.658 | 80.31 |

Table 2. Liquid state properties for the three AFM water models. The corresponding experimental properties for H2O and D2O and the TIP4P/2005 H2O values are listed for comparison. All simulated and experimental properties were for 300 K where applicable unless otherwise noted. The underlined experimental value is the most appropriate value to compare with. Bold indicates the best agreement for the property.

| | B3LYPD-4F | BLYPSP-4F | WAIL | TIP4P/2005 | Exp (H$_2$O) | Exp (D$_2$O) |
|---|---|---|---|---|---|---|
| $\rho_n$ (mol/L) | 54.49±0.02 | 54.82±0.01 | 57.31±0.02 | **55.39**[*,1] | 55.32[2] | 55.12[2] |
| $T_M$ (K, 1 bar) | 287 | 258 | **270**[3] | 252.1[4] 249[5] | 273.15[6] | 276.96[6] |
| Classical $\Delta H_{vap}$ (kcal/mol) | 12.6[7] | 11.6[8] | 11.4[9] | 10.89[*,4] | | |
| Quantum $\Delta H_{vap}$ (kcal/mol) | 11.5[7] | **10.4**[8] | 10.7[9] | | 10.51[*,6] | 10.89[*,10] |
| $D$ (10$^{-9}$ m$^2$/s) | 1.17 ± 0.02 | **2.46 ±0.08** | 1.56 ± 0.04 | 2.08[*,4] | 2.299[6] | 2.109[6] |
| $\eta$ (10$^{-3}$ kg m$^{-1}$s$^{-1}$) | 1.56±0.09 | 0.71±0.05 | 0.97±0.04 | **0.855**[*,11] | 0.853[6,12] | 1.047[6,12] |
| $\gamma$ (mN/m) | 68.5±0.9 | 62.0± 0.2 | 76.1± 0.6 | **69.3±0.9**[*,11] | 71.99[6] | 71.09[6] |
| $\varepsilon_{rv}$ | 41.85±1.96 | 42.91±1.11 | 45.38±1.18 | 60[*,4] | | |
| $\varepsilon_s$ | 74.50 ±3.50 | **76.38±1.98** | 80.77±2.10 | 106.8 | 78.408[*,6] | 78.06[+,6] |
| TMD (K, 1 atm) | 305[7] | 268[8] | **282**[9] | 278[14] | 277[6] | 284[6] |
| $T_c$ (K) | **643** | 651 | 711 | 640[15] | 647.1[6] | 643.8[6] |
| $\rho_c$ (mol/L) | **18.23** | 14.43 | 15.88 | 17.21[15] | 18.43[6] | 17.78[6] |
| $P_C$ (bar) | 72 | 115 | 125 | **146**[15] | 220.64[6] | 216.71[6] |
| $T_B$ (K) | 419 | **395** | 438 | 401[15] | 373.15[6] | 374.57[6] |

* at 298K ; + at 303 K

1, ref [7]; 2, ref [38]; 3, ref [46]; 4, ref[7]; 5, ref [42]; 6, ref [38]; 7, ref [17]; 8, ref [18]; 9, ref [19]; 10, ref[63]; 11, ref [64], 12, ref [65, 66]; 13, ref [67] 14, ref[68], 15, ref[45]

Figure Captions:

Figure 1, schematic diagram illustrating the steps in adaptive force fitting.

Figure 2, vapor pressure of TIP4P/2005 and the three AFM water models. The properties of TIP4P/2005 water is from ref [45]. The solid line without symbols is the experimental curve. The symbols with error bars indicate the measured vapor pressure; the curves were fits to the Antoine's equation that start at the $T_B$ and terminate at $T_C$.

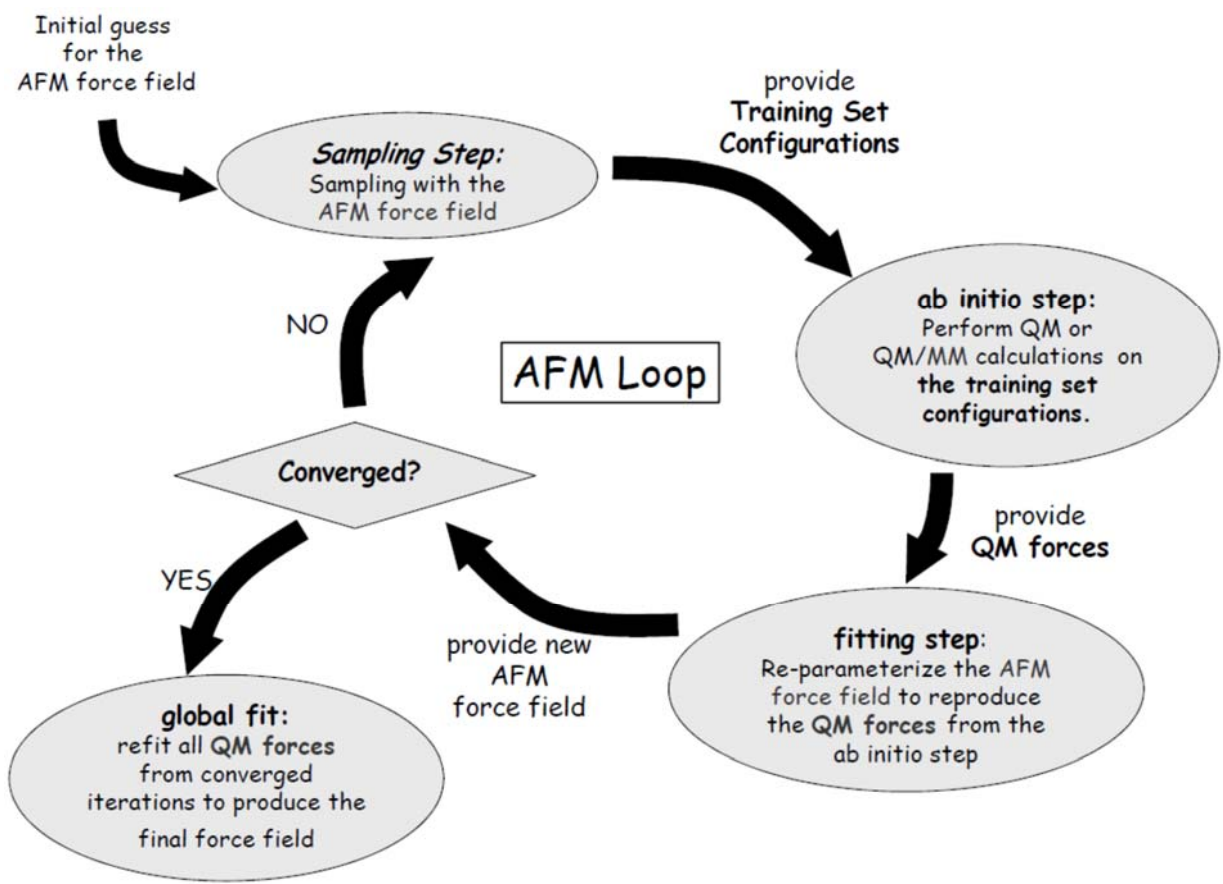

Figure 1

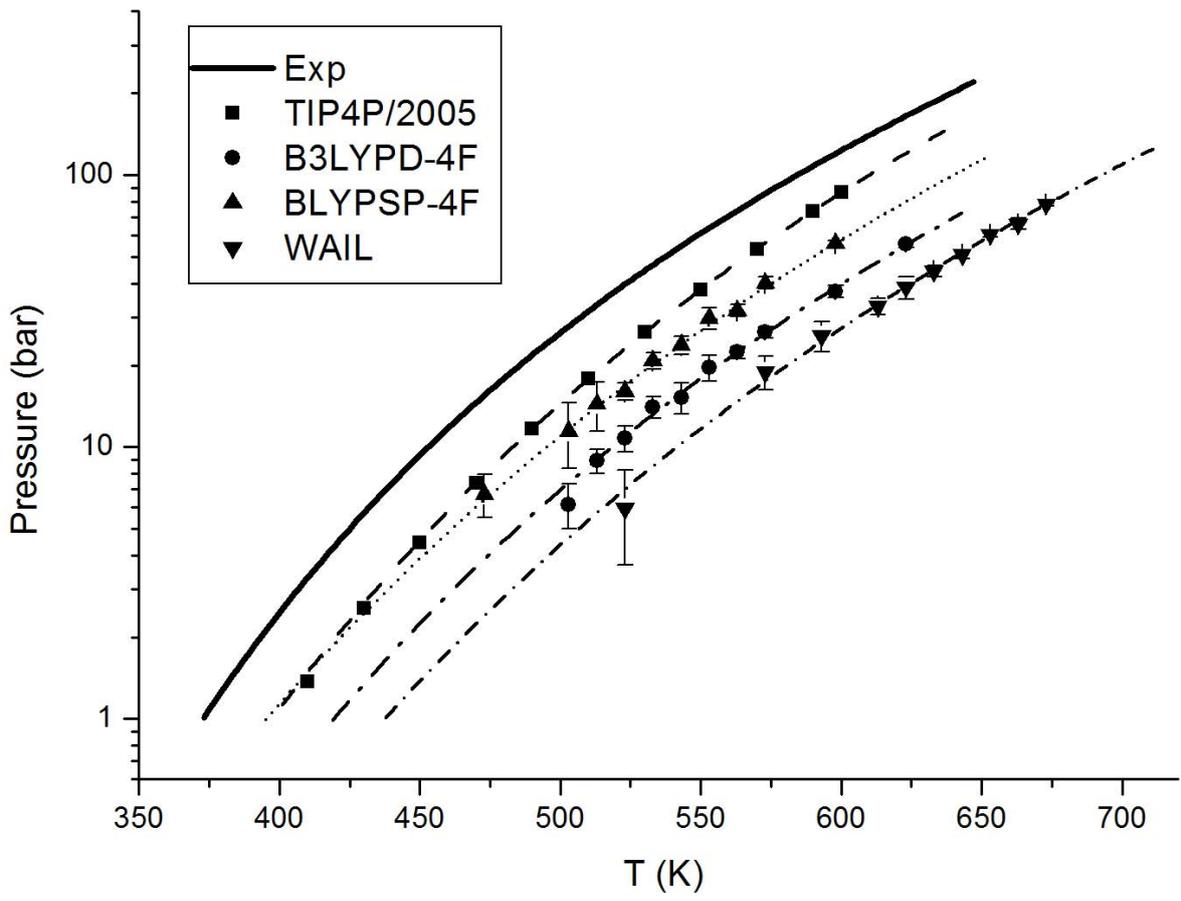

Figure 2